\documentclass[runningheads,citeauthoryear]{apinv}
\usepackage{epsfig,cite,graphics}

\usepackage[T2A]{fontenc}
\usepackage[cp1251]{inputenc}
\usepackage[bulgarian,english]{babel}

\def\Rstar {$R_\star$\,}
\def\Teff {$T_{\rm eff}$\,}
\def\vinf {$v_\infty$\,}
\def\logg {$\log g$\,}

\def\Zsun{$\rm{Z_{\odot}}$} 
\def\MV {$ M_{\rm V}$\,}

\def\vmic {$v_{\rm mic}$\,}
\begin{document}

\title{Metallicity effects in the spectral classification of O-type stars. Theoretical 
consideration.}
\titlerunning{Metallicity effects}
\author{Nevena Markova\inst{1},  Luciana Bianchi\inst{2},  Boryana Efremova\inst{2},
\and Joachim Puls\inst{3}}
\authorrunning{Markova et al.}
\tocauthor{Nevena Markova}
\institute{Institute of Astronomy, NAO, BAS,	P.O. Box 136, 4700 Smolyan, Bulgaria
	\email{nmarkova@astro.bas.bg}
\and Deptartment of Phys.\& Astron., JHU,3400 N.Charles St., Baltimore, MD21218
\and Universit\"{a}ts-Sternwarte, Scheinerstrasse 1, D-81679 M\"unchen, Germany
}
\papertype{contribution}
\maketitle
\begin{abstract}

Based on an exteded grid of NLTE, line blanketed model atmospheres with 
stellar winds as calculated by means of FASTWIND, we have investigated the
change in the strengths of strategic Helium transitions in the optical 
as caused by a 0.3 decrease in metallicity with respect to solar abundances. Our calculations 
predict that only part of the observed increase in \Teff\, of O-type dwarfs
could be explained by metallicity effects on the spectral type indicators,
while the rest must be attributed to other reasons (e.g., different stellar
structures as a function of metallicity or differences between observed and
theoretical wind parameters etc.). In addition, we found that using the
He~II~4686 line to classify stars in low metallicity environments ($Z \le$
0.3\Zsun) might artificially increase the number of low luminosity (dwarfs
and giants) O-stars, on the expense of the number of O-supergiants.
\end{abstract}

\Bg
{}
\Eng
\section*{Introduction}

Thanks to the 
outstanding work by N.~Walborn and collaborators over the last 30 years, a
detailed classification scheme for OB-stars of solar metallicity has been
developed, based on the morphology of the line spectrum. For the specific 
case of O-type stars, an alternative scheme relying on quantitative criteria
has been provided by Conti \& Alschuler \cite{CA71} and Mathys \cite{mathys}
as well. Consequently, the classification of Galactic OB stars is rigorously 
defined.

The classification of extragalactic stars, on the other hand, is still
somewhat problematic. In particular, the calibration between spectral types
and physical parameters differs as a function of metallicity when
criteria for Milky Way stars are applied. With respect to B-stars, this point
has been extensively discussed by various authors (e.g., Monteverde et al.
\cite{monteverde}, Lennon \cite{lennon97}, Urbaneja et al. \cite{urb}) who 
pointed out that the use of the MK classification criteria for stars with
metallicities different than solar can significantly influence the derived
spectral types and luminosity classes.

Unlike B-type stars, where the relative strengths of metal to He~I optical
lines are used for classification purposes, O-type stars are generally
classified by comparing the strengths of He~I and He~II lines\footnote{At
the very early O2-3 subtypes, the Helium ionization balance can no longer be
employed (due to very weak/missing He~I), and specific classification
criteria based on the strengths of optical N~V and N~IV transitions have
been developed by Walborn et al.  \cite{walborn02}.} (see, e.g., Conti \&
Alschuler \cite{CA71}; Mathys \cite{mathys}, Walborn \cite{walborn71,
walborn73}; Walborn \& Fitzpatrick,\cite{WF90}). Thus, one might conclude
that there should be no problem to employ the standard classification
schemes of either Walborn or Conti to O-type stars in various metallicity
environments. However, theoretical considerations predict that Helium line
strengths should also depend, though indirectly, on metallicity, due to the
processes of mass-loss and EUV line-blocking/blanketing (e.g., Abbott \&
Hummer \cite{AH85}, Sellmaier et al.  \cite{sellmaier93}, Herrero et al.
\cite{herrero00}). Whilst in stars with weaker winds (e.g., dwarfs) the
effect is controlled, to a major extent, by the blanketing alone, in stars
with stronger winds (e.g., supergiants), the processes of wind emission and
additional wind-blanketing seem to be of similar or even dominating impact
(e.g., Repolust et al. \cite{repo}).

\section{Metallicity effects in optical Helium line strengths}

As noted in the previous section, the classification of O-type stars in the
optical band relies either on the Walborn or on the Conti classification 
schemes, both developed to be used at solar metallicity. In the Walborn  
scheme visual estimates of individual line strengths or line strength ratios of
strategic Helium transitions are used as classification criteria, while 
in the Conti scheme the logarithm of the equivalent width (EW) ratio of He~I~4471
to He~II~4541 is used instead. Given that an eye-estimated ratio is roughly
equivalent to the logarithmic ratio of the corresponding EWs, these two
approaches should lead to similar results (but see Heap et al.
\cite{heap06}).

To address the metallicity issue regarding the spectral classification of
O-type stars, we developed the following strategy.  As a first step and
using the FASTWIND code (Puls et al. \cite{puls05}), we constructed 6 model
grids, corresponding to two values of metallicity, solar, $Z_{\odot}$, and
0.3~$Z_{\odot}$\footnote{This investigation was designed and carried out as
a first step of a project to study the massive stellar content in NGC~6822,
for which a mean iron abundance, [Fe/H], of about -0.5 was found,
corresponding to an average metallicity of about 0.3~\Zsun (for more
information, see Efremova et al.  \cite{efremova} and references therein).},
and three luminosity classes: dwarfs (DWs), giants (Gs) and supergiants
(SGs). By means of these models we investigated the predicted 
variations in Helium line strengths, caused by the adopted change in 
metallicity, as a function
of \Teff\ and luminosity class. Subsequently, we evaluated the shifts in
spectral type (and \Teff) and luminosity class, that would eventually appear
if Galactic classification criteria were used to O-type stars of metallicity
0.3~\Zsun.

\subsection{Model grids}

Our model grids comprise 64 NLTE, line-blanketed models with stellar winds.
For all models and independent of metallicity,  stellar parameters (\Teff, 
\logg, and \Rstar) were taken from the empirical calibrations of Martins 
et al.  \cite{martins05} for Galactic stars of luminosity classes I, 
III and V. We used a solar chemical composition as derived by Asplund et 
al.\cite{asplund}, subsequently scaled to Z=0.3~$Z_\odot$.

A standard $\beta$-velocity law with $\beta$=0.9 was adopted for all models.
Terminal velocities, \vinf, for solar metallicity models were determined by
means of the spectral type - \vinf calibration from Kudritzki \& Puls
\cite{KP}. The effects of metallicity on \vinf were taken into account
by employing the scaling relation from Leitherer et al. \cite{leitherer},
namely \vinf $Z^{\bf 0.13}$. Mass-loss rates were calculated using the
mass-loss recipe from Vink et al. \cite{vink01}. A microturbulent
velocity of 10~\kms was used to calculate the atmospheric structures, whilst 
values of 5, 10 and 15~\kms were assumed for calculating the emergent
spectrum. 

\subsection{Temperature classification}

Based on these model grids, the EW variations of several strategic He~I (e.g.,
He~I~4026, He~I~4471 and He~I~4387) and He~II (e.g., He~II~4200, He~II~4541
and He~II~4686) lines were investigated, as a function of metallicity (\Zsun\
and 0.3~\Zsun), \Teff\, and luminosity class.
\begin{figure}[t]
\begin{minipage}{6.8cm}
\resizebox{\hsize}{!}
{\includegraphics{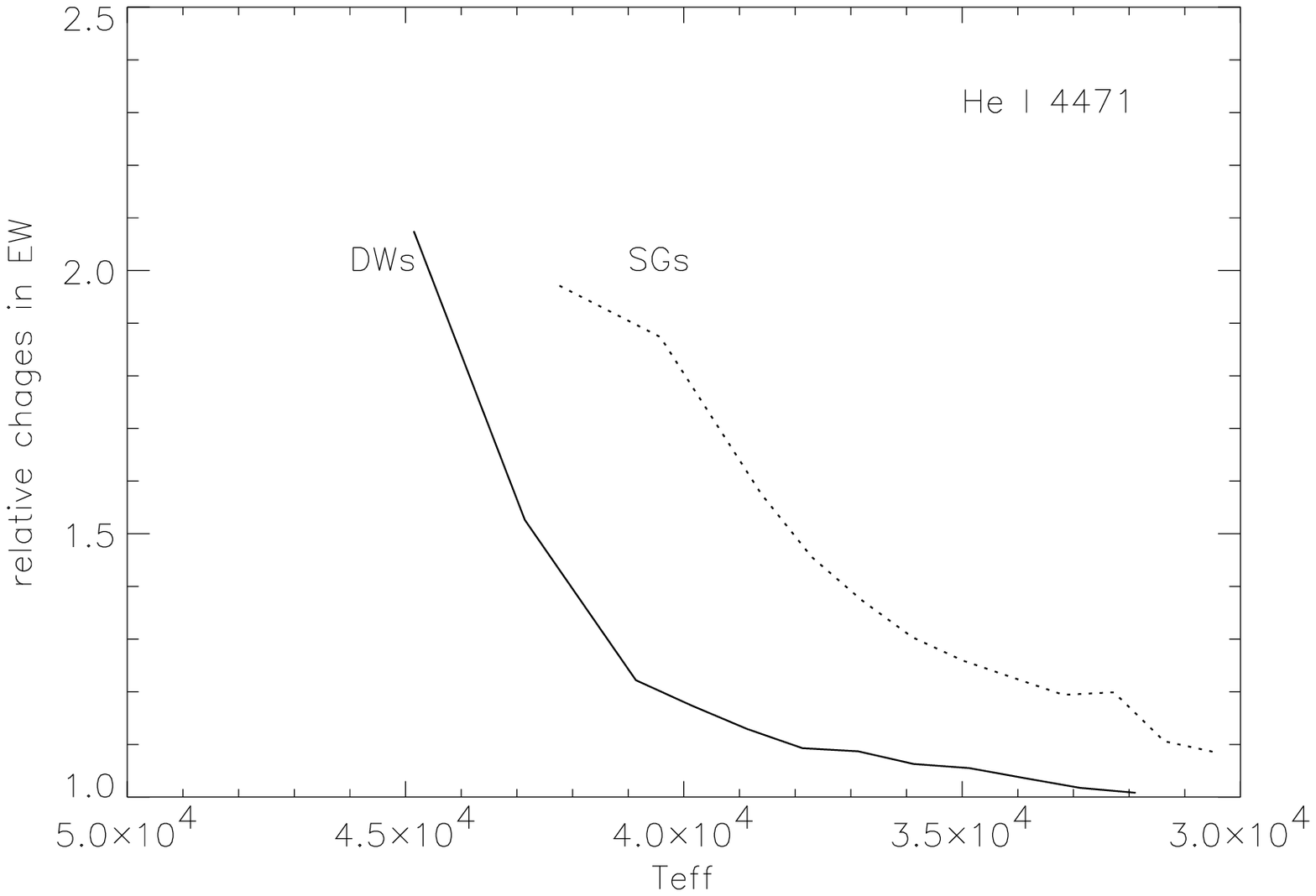}}
\end{minipage}
\hfill
\begin{minipage}{6.8cm}
\resizebox{\hsize}{!}
{\includegraphics{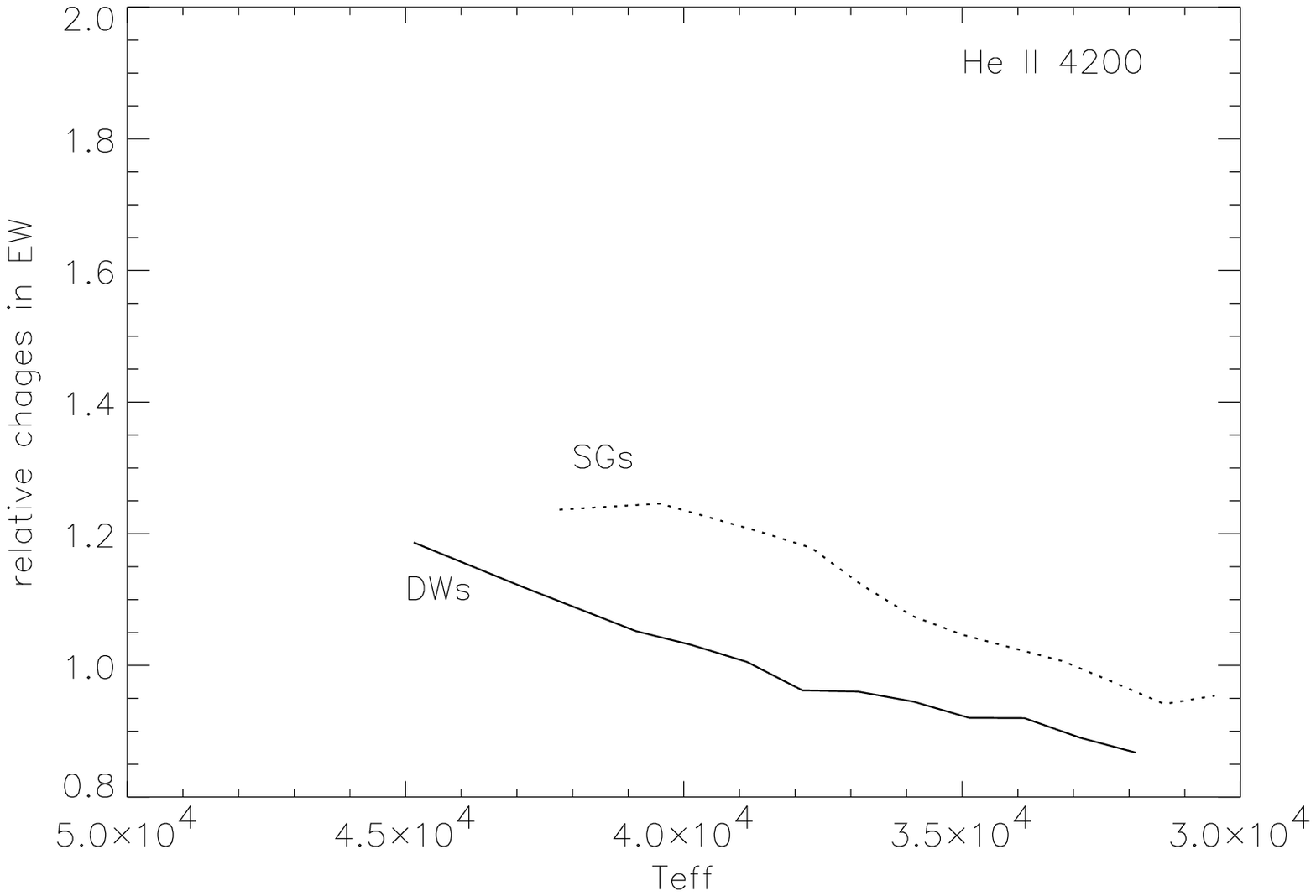}}
\end{minipage}
    \caption[]{Exemplary changes in Helium line strengths (in units
of EW(Z = 0.3~\Zsun)/EW(\Zsun)) as caused  by the 
adopted decrease in metallicity.  
}
\label{he_lines}
\end{figure}
Our results, partly illustrated  in Figures~\ref{he_lines} 
and 3, indicate that in the O-star temperature domain He~I and 
He~II are both influenced by metallicity. The effect is  
opposite to that for metal lines, i.e., while metal lines 
become weaker due to the lower metal content, the Helium line 
strengths are predicted to increase in absorption.

Moreover, individual lines react differently. Members of the
He~I singlet series are more sensitive to metallicity than those of the
triplet series\footnote{This fact has been pointed out by Herrero et al.
\cite{herrero00}; see also Najarro et al. \cite{najarro} for problems in
modeling the singlet lines.}, which in turn are more sensitive than He~II
transitions (except for He~II~4686, see next section).  In addition, the
effect depends on \Teff\, and \logg, being more pronounced in hotter SGs
(where mass loss effects are strongest), and almost negligible (within
$\pm$10\%)\footnote{For good quality spectra the error of individual
EW measurements is typically of the order of about 10 percent.} at intermediate and cooler
temperature DWs. 

Our findings imply that, due to the lower metal content, the optical Helium
line spectrum of low metallicity O-stars would appear somewhat different
from that of their Galactic conterparts. In particular, at Z=0.3~\Zsun\, and
if the strength of the primary classification lines, He~I~4471 and
He~I~4387, were to be considered, these lines would resemble those of a
Galactic star with {\it lower} \Teff\ (assuming the same \logg\ and \Rstar): 
about 2~kK lower for SGs and up to 1~kK lower for DWs. Vice versa, if compared to 
Galactic counterparts with the {\it same observed} EWs in He~I,
stars with a lower metal content should be hotter.
 
For the strategic He~II~4200 and He~II~4541 transitions we find the
following situation: for intermediate and low temperature DWs (39 $\le$
\Teff $\le$ 32 kK) and cooler SGs (\Teff\ below 34kK), the corresponding lines 
strengths are predicted to remain rather unaffected by metallicity, whilst at
higher \Teff\ they are amplified by up to 20 percent, mimicking the lines of
a Galactic star of {\it higher} \Teff. 
\begin{figure}[t]
\begin{minipage}{6.8cm}
\resizebox{\hsize}{!}
{\includegraphics{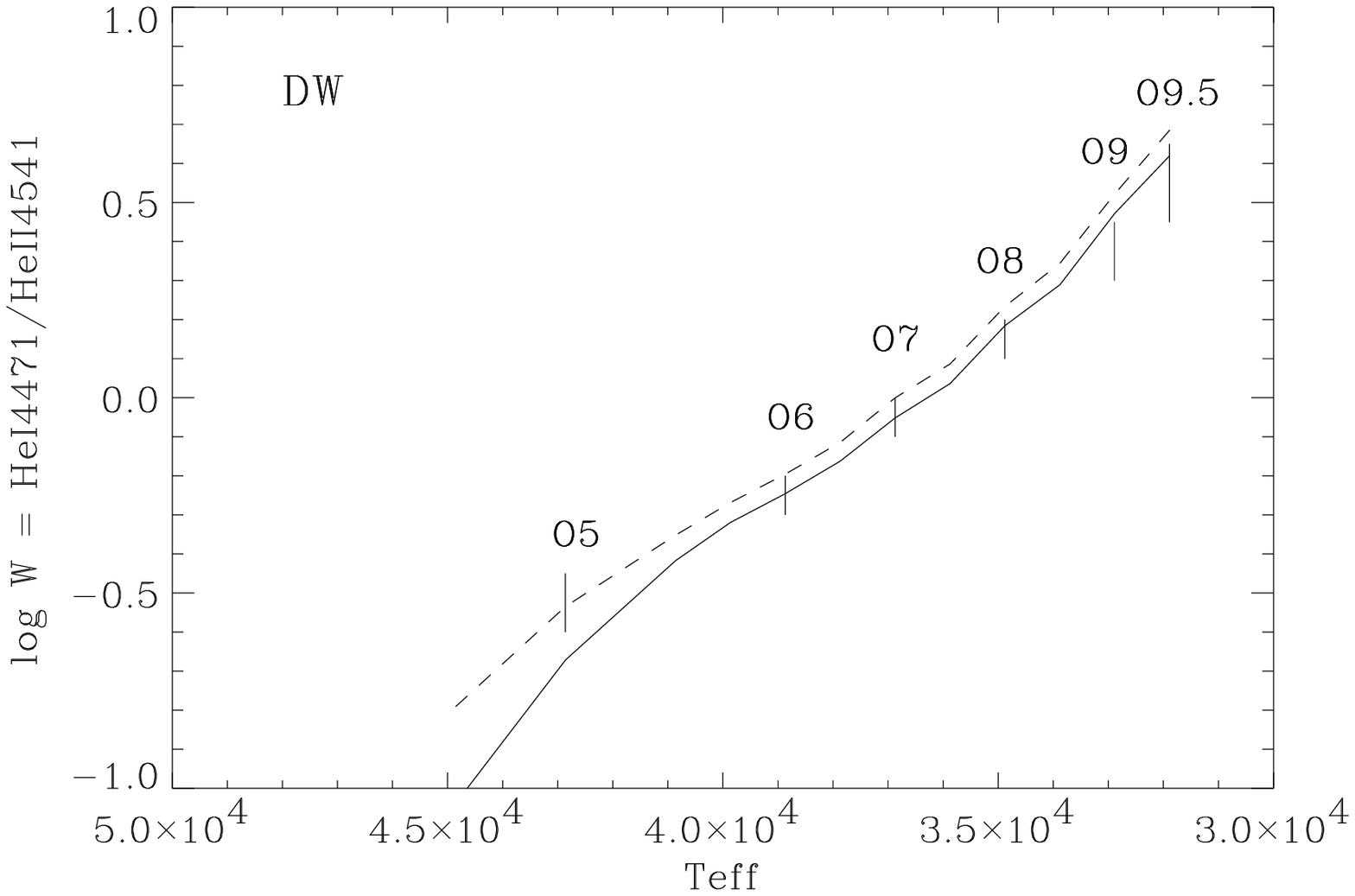}}
\end{minipage}
\hfill
\begin{minipage}{6.8cm}
\resizebox{\hsize}{!}
{\includegraphics{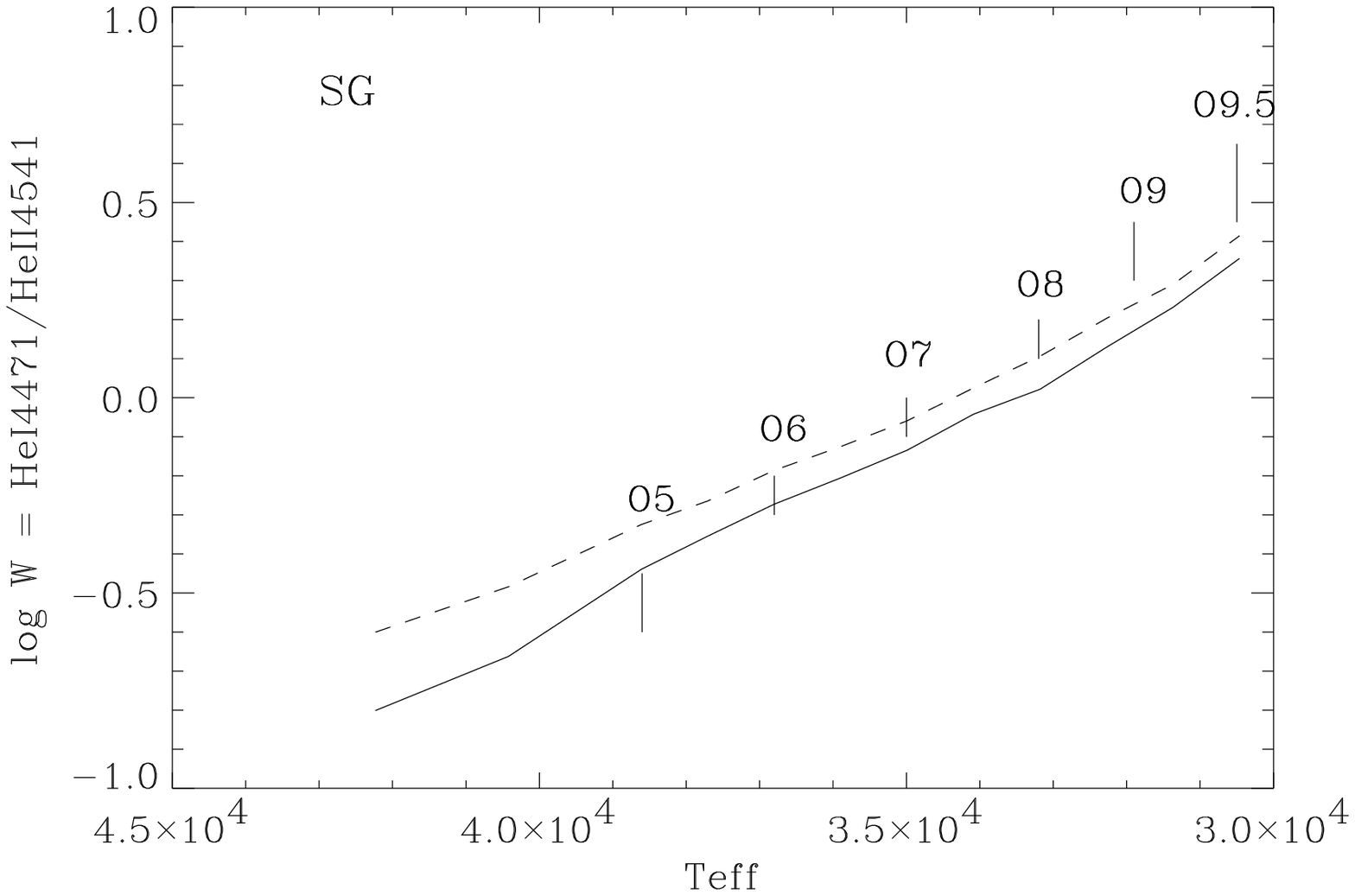}}
\end{minipage}
    \caption[]{Predicted He~I~4471/He~II~4541 line 
ratios  for Z=\Zsun\, (solid) and Z=0.3~\Zsun\, (dashed). Vertical  
lines mark the range of values determining each spectral type  
according to the Mathys  $\log W$ - spectral type calibration.
}
\label{he_ratio}
\end{figure}

The effect of metallicity on the spectral classification of O-stars can be
evaluated by investigating the behaviour of He~I/He~II line ratios instead
of individual line strengths.  Thus, we have calculated and compared the
logarithmic He~I(+II)~4026/He~II~4200, He~I~4471/He~II~4541 and
He~I~4387/He~II~4541 line ratios, as a function of \Teff, for the cases of
DWs, Gs and SGs and for the two values of metallicity, \Zsun\, and
0.3~\Zsun. The results of these calculations are partly illustrated in 
Figure~\ref{he_ratio}, where the lower and upper limits of the line 
strength ratios from the Mathys $\log W$ - spectral type
calibration\footnote{The Conti classification scheme has been extended by
Mathys \cite{mathys} to cover the earliest and latest O-subtypes.} are
overplotted as vertical lines. The x-position of each of these lines has
been placed at the appropriate \Teff-value following from the Martins et al.
empirical calibration for the corresponding spectral type.

Apparently, this criterion to classify O-stars with Z=0.3~\Zsun\, would lead
to temperature differences that increase with \Teff\ and luminosity class. In
particular, for SGs these differences are predicted to range from about 1 
to about 2 kK (i.e. about one sub-type), while for DWs the expected values
are a factor of 2 smaller. 

Another important result to be noted is that whilst for Galactic DWs 
a good correspondence between the calculated and the calibrated 
values of the He~I~4471/He~II~4541 line ratios is found, 
for cooler SGs (subtypes later than O7) the calculated values  are
systematically smaller. This discrepancy results from the (still
not understood) so-called ``generalized dilution effect'', which makes the
synthetic He~I~4471 lines in low temperature SG models weaker than
observed (see Repolust et al. \cite{repo} and references therein). Since
this effect seems to be independent on metallicity, it is not expected
to influence our general findings though.

\subsection{Luminosity classification}

In the Walborn classification scheme, the main luminosity criterion is He~II~4686. 
In SGs this line appears only in emission, while in DWs  it is observed 
as a pure absorption feature. Since the amount of emission depends on 
the wind properties, which in turn depend on metallicity, one expects 
the strength of He~II~4686 to depend on metallicity as well (see, e.g., 
Massey et al. \cite{massey04}). 

In Figure~3 the behaviour of the He~II~4686 EWs (positive for emission and
negative for absorption) are shown as a function of \Teff, 
for two luminosity classes (SGs and DWs) and for Z=\Zsun\ and 0.3~\Zsun.
Expectedly, our calculations for solar metallicity models produce He~II~4686
in emission for SGs and in absorption for DWs.  Interestingly, however, our 
calculations predict that the metallicity effects on this line can be so
strong that at Z=0.3~\Zsun\, its sensitivity to luminosity vanishes almost
completely, and He~II~4686 appears in absorption for all calculated models. 
\begin{figure}[t]
\begin{centering}
\begin{minipage}{6.8cm}
\resizebox{\hsize}{!}
{\includegraphics{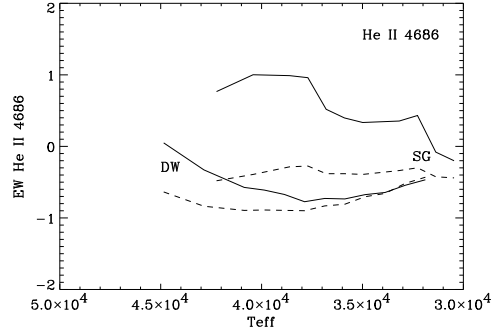}}
\end{minipage}
    \caption[]{Equivalent widths (positive for emission)
    of He~II~4686 as a function of \Teff, calculated for two values  
of metallicity - \Zsun (solid) and 0.3~\Zsun (dashed), and for two luminosity 
classes (SGs and DWs). 
}
\end{centering}
\label{heii4686}
\end{figure}

Finally let us note that although an increase in microturbulent 
velocity (from 5 to 15 \kms) is predicted to shift the assigned 
spectral types by half (for DWs)\footnote{ This estimate is 
consistent with Mokiem et al. \cite{mokiem04} who found a similar 
result for the case of a hot (O5) dwarf.} to one (for SGs) subtype 
towards later spectral types, this effect does not seem to depend on metallicity. Due to this reason model predictions  corresponding 
to \vmic = 10~\kms\, only  are shown throughout this paper.

\section{Discussion and Conclusions}

Based on a grid of NLTE, line blanketed model atmospheres with 
stellar winds as calculated by means of FASTWIND, we found that

\noindent $\bullet$ in O-stars the optical Helium spectrum 
can be significantly influenced by metallicity effects, where 
a decrease in metal content results in stronger He line absorptions 
{\it from both ionization stages}, where lines from He~I are more 
severely affected than those from He~II. Individual lines react 
differently in dependence of \Teff\ and luminosity class, 
but do not depend on microturbulence. This result is
particularly important when {\it individual line strengths} are used to
classify stars, as it is the case for the Walborn scheme where the detection
of weak He~I~4471 and He~I~4387 is used to fix the subtypes O4 and O6,
respectively. According to our calculations at $Z$=0.3~\Zsun, the use of
these lines as classification indicators would shift the assigned spectral
types by up to one subtype to the later side, provided the detection
threshold is the same.

\noindent
$\bullet$ 
Furthermore, our calculations showed that not only individual 
Helium lines, but Helium line ratios are also varying with
metallicity. In particular, the use of the 
He~I~4471/He~II~4541 ratio to classify O-stars of $Z$=0.3~\Zsun\, 
would shift the derived temperature scale  
by up to 2~kK (i.e., about one sub-type) for SGs,  and 
up to 1~kK (i.e., about half a sub-type) for DWs to 
the hotter side.  The latter result 
is consistent  with the predictions by Mokiem et al. 
\cite{mokiem04}, who found, using CMFGEN (Hillier \& Miller 
\cite{hillier}), that variations in the metal content from Z=2~\Zsun\, 
to Z=0.1~\Zsun\, can  change the spectral type of a hot DW by up 
to one and a half sub-type.

Extended extragalactic surveys of hot massive stars showed that 
O-stars in low metallicity environment are systematically hotter 
than their Galactic counterparts with the same spectral type 
(e.g., Mokiem et al. \cite{mokiem06}, Massey et al. 
\cite{massey04, massey05}, Efremova et al. 2009), thus
confirming our principal predictions. However, the observed shifts 
in \Teff\ are significantly larger than those predicted to appear 
from a shift in the assigned spectral types, as calculated here and 
in other work. In particular, for O-type DWs temperature differences of 
3 to 4 kK, i.e., a factor of 3 to 4 larger than those predicted by 
Mokiem et al. \cite{mokiem04} were established between objects 
in our Galaxy and in the SMC (Z=0.2~\Zsun)\footnote{According to the 
computations by Mokiem et al., a factor of 5 decrease in $Z/$\Zsun\ 
would result in a shift of only 1~kK in \Teff\ of a hot DW, i.e., about 
half a subtype.}. The corresponding  results for DWs in  NGC~6822 are 
roughly the same (Efremova et al. \cite{efremova}). 

Thus, only part ($\sim$25 to 30 percent) of the observed differences in 
\Teff\ of O-type DWs in the SMC and in NGC~6822 could be explained by metallicity effects on the spectral type indicators, while the rest 
must be attributed to other, more ``physical''reasons, such as, e.g., different stellar structures as a function of metallicity
or differences between observed and theoretical wind parameters etc.
Additional calculations are required to separate and evaluate these 
effects.

\noindent $\bullet$ Our calculations predict that the amount of wind
emission in He~II~4686 diminishes towards lower metallicities,
so that at $Z \le$0.3~\Zsun\, this line would appear only in absorption and could
no longer serve as a reliable luminosity indicator. In particular, using
this line to classify stars in low metallicity environments might
artificially increase the number of low luminosity (DWs and Gs) O-stars
identified in  galaxies with lower metallicity, on the expense of 
the number of O-SGs.

Some hints about such effects seem to have been established already. For
instance, Massey et al. \cite{massey04} recently noted that in several of
their MC targets the He~II~4686 emission is weaker than it might be
expected, given their absolute magnitudes, \MV. And although one might argue
that a discrepantly high \MV\, could be due to spectroscopic binarity, an
alternative explanation in terms of metallicity effects is also possible,
accounting for our present results. Moreover, the number of SGs identified
in the SMC appears to be less than expected (D. Lennon, priv. comm.), in
agreement with what might be expected from our calculations.

{\it Acknowledgments:} Part of this work was performed during a visit of NM
at the dept. of Physics and Astronomy of the Johns Hopkins University.
The work was supported by NASA grant NAG5-9219 (NRA-99-01-LTSA-029, PI Bianchi).
J.P. acknowledges a travel grant by the DFG (under grant Pu 117/6-1) and 
by the BG NSF (under grant F-1407/2004).

\end{document}